%% file: main.tex
\newif\ifarxiv
\arxivtrue

\newcommand{\arxivonly}[1]{%
  \ifarxiv
    #1%
  \fi
}
\newcommand{\conferenceonly}[1]{%
  \ifarxiv
  \else
    #1%
  \fi
}

\ifarxiv
  \documentclass[sigconf,nonacm]{acmart}
\else
  \documentclass[sigconf,review]{acmart}
\fi

\AtBeginDocument{%
  }

\input{sections/macro}

\usepackage{multirow}

\usepackage{graphicx}
\usepackage{subcaption}
\usepackage{mathtools}
\usepackage[most]{tcolorbox}

\newtcolorbox{FindingBox}[1]{
  enhanced,
  breakable,
  colback=cyan!3!white,
  colframe=cyan!45!blue!60,
  boxrule=0.5pt,
  arc=2.5pt,
  boxsep=2pt,
  fonttitle=\bfseries,
  title=#1,
  boxed title style={
    colback=cyan!8!white,
    colframe=cyan!45!blue!60,
    boxrule=0pt,
    arc=2pt,
    left=4pt,right=4pt,top=2pt,bottom=2pt
  }
}

\setcopyright{none}
\renewcommand\footnotetextcopyrightpermission[1]{} 
\newcommand{\modelname}{EST}

\newcommand{\acronym}[1]{\underline{\textbf{#1}}}

\makeatletter
\def\blfootnote{\gdef\@thefnmark{}\@footnotetext}
\makeatother

\usepackage{cleveref}

\begin{document}

\title{EST: Towards Efficient Scaling Laws in \\ Click-Through Rate Prediction via Unified Modeling}

\author{Mingyang Liu$^{*}$\ali, Yong Bai$^{*}$\ali, Zhangming Chan$^{\ddagger}$\ali, Sishuo Chen\ali \\ Xiang-Rong Sheng\ali, Han Zhu\ali, Jian Xu\ali $^{\dagger}$, 
    Xinyang Chen$^{\dagger}$} 
    \email{{mingyangliu1024, chenxinyang95}@gmail.com} 
    \email{ {baiyong.by,zhangming.czm,chensishuo.css,xiangrong.sxr,zhuhan.zh,xiyu.xj}@alibaba-inc.com}
    \affiliation{
        \institution{\ali Taobao \& Tmall Group of Alibaba, Beijing, China}
        \country{}
}

\renewcommand{\shortauthors}{Liu et al.}

\input{sections/abstract}

\conferenceonly{}


\maketitle
\input{sections/introduction2}
\arxivonly{
    \input{sections/related}

}

\input{sections/preliminaries}

\input{sections/method}
\input{sections/experiment}
\input{sections/conclusion}
\arxivonly{
    \input{sections/acknowledge}
}

\balance
\bibliographystyle{ACM-Reference-Format}
\bibliography{ref}

\conferenceonly{
    \newpage
    \appendix
    \input{sections/appendix}
    \label{sup}
}

\end{document}

%% file: sections/macro.tex
\usepackage{enumitem}
\usepackage[dvipsnames]{xcolor}
\usepackage{amsmath}
\usepackage{xfrac}
\usepackage[ruled,vlined,linesnumbered]{algorithm2e}
\usepackage{microtype}
\usepackage{balance}

\definecolor{mybluegray}{RGB}{89, 132, 132} 

\SetCommentSty{mycommfont}

\let\oldnl\nl
\newcommand{\nonl}{\renewcommand{\nl}{\let\nl\oldnl}}

\definecolor{DSgray}{cmyk}{0,1,0,0}

\def \ali {$^{\hspace{.1em}{\color{purple}\boldsymbol{a}}}$}

%% file: sections/abstract.tex

\begin{abstract}
Efficiently scaling industrial Click-Through Rate (CTR) prediction has recently attracted significant research attention. Existing approaches typically employ early aggregation of user behaviors to maintain efficiency. However, such \textit{non-unified} or \textit{partially unified} modeling creates an information bottleneck by discarding fine-grained, token-level signals essential for unlocking scaling gains. In this work, we revisit the fundamental distinctions between CTR prediction and Large Language Models (LLMs), identifying two critical properties: the \textit{asymmetry in information density} between behavioral and non-behavioral features, and the \textit{modality-specific priors} of content-rich signals. Accordingly, we propose the Efficiently Scalable Transformer (EST), which achieves fully unified modeling by processing all raw inputs in a single sequence without lossy aggregation. EST integrates two modules: Lightweight Cross-Attention (LCA), which prunes redundant self-interactions to focus on high-impact cross-feature dependencies, and Content Sparse Attention (CSA), which utilizes content similarity to dynamically select high-signal behaviors. Extensive experiments show that EST exhibits a stable and efficient power-law scaling relationship, enabling predictable performance gains with model scale. Deployed on Taobao's display advertising platform, EST significantly outperforms production baselines, delivering a \textbf{3.27\%} RPM (Revenue Per Mile) increase and a \textbf{1.22\%} CTR lift, establishing a practical pathway for scalable industrial CTR prediction models.

\blfootnote{$^*$Equal contribution. $^\dagger$Corresponding Author. 
\arxivonly{
    \\
    $^\ddagger$This author also made core contributions to methodology of this work.
}
}
\end{abstract}

%% file: sections/introduction2.tex
\begin{figure}[htbp]
    \centering
    \includegraphics[width=\linewidth]{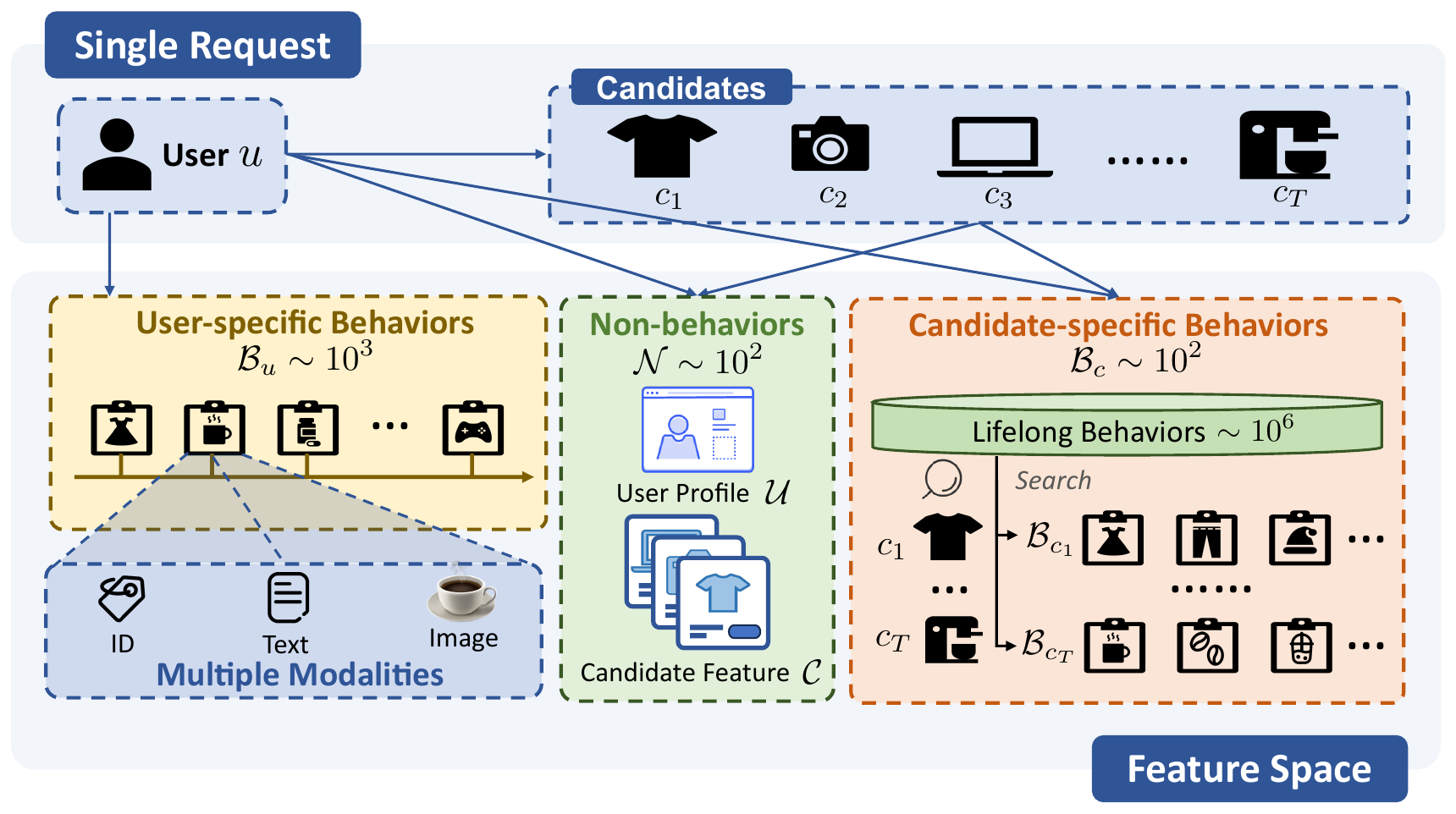}
    \vspace{-15pt}
    \caption{The CTR prediction in recommendation systems.}
    \vspace{-10pt}
    \label{fig: problem_setting}
\end{figure}

\section{Introduction}

\begin{figure*}[!t]
    \centering
    \includegraphics[width=\textwidth]{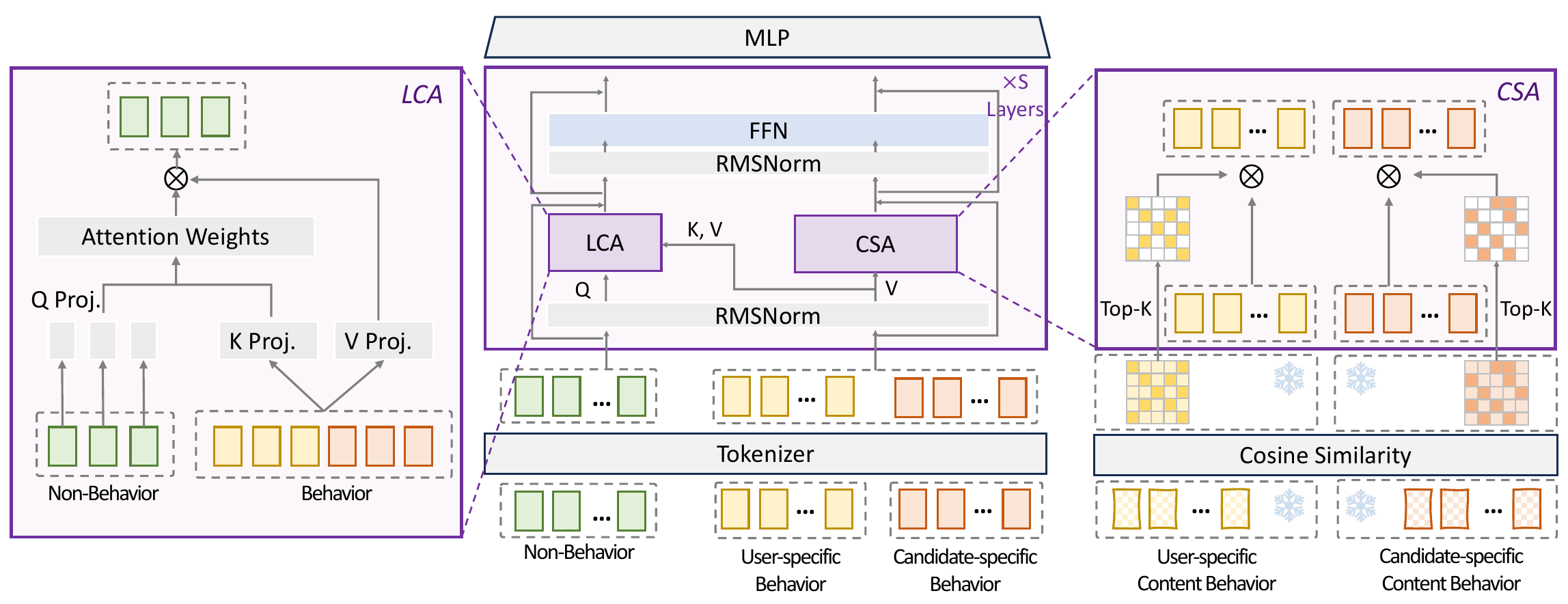}
    \caption{The architecture of {\modelname}.
    Feature-specific tokenizers map raw features to corresponding tokens. Lightweight Cross-Attention (LCA) captures the interactions between non-behavioral tokens and behavioral tokens. Content Sparse Attention (CSA) calculates intra-sequence content similarity and performs sparse attention within the behavior sequences.}
    \label{fig: model_architecture}
\end{figure*}

Click-through rate (CTR) prediction is a fundamental task in modern recommendation systems, aimed at estimating the probability of a user clicking on an item. As illustrated in~\Cref{fig: problem_setting}, industrial CTR prediction models typically integrate diverse inputs, including non-behavioral features $\mathcal{N}$ (e.g., user profiles, candidate attributes), user-specific behaviors $\mathcal{B}_{u}$ (e.g., short-term historical behaviors) and candidate-specific behaviors $\mathcal{B}_{c}$ (e.g., subsequences dynamically searched from lifelong behaviors with the candidate item~\citep{pi2020search,cao2022sampling,chang2023twin,wu2025muse}). Moreover, these behaviors are often multimodal~\cite{sheng2024enhancing}, interleaving discrete IDs with rich content signals such as images and text. These features are fed into deep networks to model feature interactions and produce final predictions.

Recently, inspired by the transformative success of scaling laws~\citep{kaplan2020scaling,henighan2020scaling,hoffmann2022training,achiam2023gpt,yang2025qwen3,liu2024deepseek} in large language models (LLMs), researchers have actively pursued analogous scaling strategies for CTR prediction~\citep{guo2017deepfm,shan2016deep,zhou2018deep,song2019autoint,wang2021dcn,xia2023transact}. 
However, industrial recommendation systems impose strict practical constraints: models must score thousands of candidates per user request within millisecond-level latency budgets.
This requirement creates a major bottleneck for standard transformer architectures, whose computational cost scales with both model size and input sequence length.
In industrial CTR prediction settings, the unified input sequence, which comprises user-specific behaviors $\mathcal{B}_u$, non-behavioral features $\mathcal{N}$, and candidate-specific behaviors $\mathcal{B}_c$, can reach a scale of $10^3$ tokens. 
The stringent high-throughput and low-latency requirements necessitate the investigation of \textbf{efficient scaling laws}, aimed at maximizing scaling gains under a constrained compute budget.

To achieve this goal, existing solutions fall into two paradigms\conferenceonly{\footnote{More detailed discussion about related works is provided in Appendix~\ref{sup}.}}:
\begin{itemize}[leftmargin=*]
    \item \emph{Hierarchical modeling}~\citep{zhang2024wukong,zhu2025rankmixer,gui2023hiformer,yan2025scaling,yu2025hhft,li2025infnet} aims to maintain efficiency by compressing behavioral sequences ($\mathcal{B}_u$ and $\mathcal{B}_c$) into fixed-length representations.
    This compression is typically achieved via modules like DIN \citep{zhou2019deep} or LONGER \citep{chai2025longer} before joint modeling with non-behavioral features $\mathcal{N}$. Though efficient, this \emph{early aggregation} creates a significant information bottleneck, discarding fine-grained token-level signals in behaviors that are essential for unlocking scaling gains.
    \item \emph{Unified modeling}~\citep{zhang2025onetrans,han2025mtgr} attempts to preserve these signals by jointly modeling $\mathcal{B}_u$ and $\mathcal{N}$ within a single sequence and leverages computation reuse to enable deeper interaction modeling under compute budget, \textit{i.e.}, sharing user-side calculations across multiple candidates in the same request. 
    Nevertheless, existing unified approaches remain \emph{partially unified}: candidate-specific behaviors $\mathcal{B}_c$ cannot be shared across candidates and are often pooled into fixed-length summaries, causing information loss.
\end{itemize}

These limitations suggest that the effective scaling of CTR  prediction models necessitates architectures that leverage domain-specific \textit{inductive biases}. In this work, we analyze the foundational distinctions between CTR prediction and LLM modeling to design an efficiently scalable architecture that achieves \emph{fully unified modeling}. We identify two critical insights: (1) \textbf{Asymmetry in Information Density}. Unlike the relatively homogeneous token sequences in LLMs, CTR inputs exhibit an asymmetry: concise, high-signal non-behavioral features $\mathcal{N}$ (the \textit{query}) must interact with massive, low-density behavioral sequences $\mathcal{B}$ (the \textit{context}). This structural prior makes dense all-to-all interactions, such as standard self-attention, computationally redundant under strict compute budget, as most behavior-to-behavior interactions contribute marginal value to the final prediction. (2) \textbf{Modality-specific Priors}. User behaviors are structured records interleaving multiple modalities, specifically discrete IDs and dense content signals. These heterogeneous signals require different utilization strategies. Rather than naively treating content features as homogeneous token embeddings, they are often more effective when utilized as \textbf{similarity-guided priors}~\citep{yang2023courier,sheng2024enhancing,wu2025muse} to bridge the semantic gap and guide the interaction within the ID space.

Building on these properties, we conduct a detailed empirical analysis that yields two actionable insights:
\begin{enumerate}[leftmargin=*]
    \item \textbf{Information density guides interaction priority.} Our analysis reveals that interactions between features of varying information densities yield disproportionate returns. Specifically, cross-attention between non-behavioral features $\mathcal{N}$ (as \textit{queries}) and behavioral sequences $\mathcal{B}$ (as \textit{keys/values}) is the primary driver of performance, whereas behavior-to-behavior interactions are largely redundant. This motivates \textbf{Lightweight Cross-Attention (LCA)}, which prunes low-gain self-interactions and retains only the critical cross-interactions, effectively decoupling model scaling from sequence length.
    
    \item \textbf{Content signals are most effective as relational priors.}  We find that content signals (e.g., images and text) deliver the largest gains when utilized to compute similarity relations rather than being directly injected as token embeddings. Accordingly, we propose \textbf{Content Sparse Attention (CSA)}, which leverages content-based similarity to dynamically select the most relevant behaviors. By performing attention only on this high-signal subset, CSA achieves superior predictive performance with nearly negligible computational overhead.
\end{enumerate}

Overall, we propose \textbf{EST}, an \textbf{E}fficiently \textbf{S}calable \textbf{T}ransformer-based architecture. As shown in~\Cref{fig: model_architecture}, in \textbf{EST}, all raw inputs---comprising non-behavioral features $\mathcal{N}$, user and candidate behaviors ($\mathcal{B}_u$ and $\mathcal{B}_c$), and their associated content signals---are organized into a \textbf{fully unified sequence} without lossy prior aggregation. This representation is processed by a scalable backbone where each layer computes the proposed LCA and CSA in parallel, enabling the model to effectively integrate heterogeneous tokens within a single framework. Experimental results demonstrate that \textbf{EST} exhibits a highly efficient power-law scaling relationship: predictive performance improves consistently as model parameters and compute budget increase, significantly outperforming existing hierarchical and partially unified paradigms.

In summary, the major contributions of this work are threefold:
\begin{itemize}[leftmargin=*]
    \item \textbf{Foundational Insights}: We identify the foundational distinctions between CTR prediction and LLMs, specifically \textit{asymmetry in information density} and \textit{modality-specific priors}. From these, we derive actionable inductive biases that guide the design of efficiently scalable architectures for recommendation.
    \item \textbf{Efficiently Scalable Architecture}: We propose \textbf{EST}, a novel architecture that achieves completely unified modeling of all raw inputs within a single sequence, eliminating the~need for lossy early aggregation. \textbf{EST} consists of \textbf{LCA} and \textbf{CSA}, which drastically reduce computational redundancy by prioritizing high-signal interactions, enabling the model to decouple its scaling potential from the prohibitive overhead of long sequences.
    \item \textbf{Industrial Deployment and Practical Insights}: Extensive offline and online experiments within Taobao's display advertising system demonstrate that \textbf{EST} exhibits a stable and efficient power-law scaling relationship. These findings and the practical deployment experience provide a valuable roadmap for practitioners seeking to explore efficient scaling in industrial CTR prediction.
\end{itemize}

%% file: sections/related.tex
\section{Related Work}
\label{Related Work}

\noindent\textbf{Deep CTR Prediction Models.}~With the advancement of deep learning, CTR prediction models have evolved from traditional machine learning~\citep{richardson2007predicting,juan2016field} to neural networks~\citep{guo2017deepfm,shan2016deep,wu2022adversarial,zhang2022keep,hu2023ps}, typically comprising sparse lookup tables that map ID features to embeddings and a dense network that fuses these embeddings for final prediction. 
Within the dense component, modeling user behavior to capture user interests is a key paradigm. Early approaches~\citep{hidasi2015session,zhai2016deepintent} use RNNs to model behavior sequences, while later models~\citep{zhou2019deep,chai2025longer,pi2020search,cao2022sampling,chang2023twin,wu2025muse} aggregate behaviors  by computing target attention with candidate items.
Another major direction focuses on high-order feature interaction. Methods such as gate-based networks~\citep{wang2021dcn,huang2019fibinet,wang2021masknet,bian2020can} and attention mechanisms~\citep{song2019autoint,xia2023transact,li2025infnet,zeng2025interformer} have been widely adopted for feature interaction. However, many such architectures remain limited in parameter scale and computational efficiency, hindering their scalability and generalization.

\vspace{\baselineskip} 

\noindent\textbf{Scalable CTR Prediction Models.}~Inspired by the scaling law~\citep{kaplan2020scaling,henighan2020scaling,hoffmann2022training} observed in large language models, recent studies have sought to scale up the dense components of CTR prediction models, revealing a substantial potential for performance improvement. Initial efforts primarily focused on scaling the feature interaction module. For instance, WuKong~\citep{zhang2024wukong} employs a Factorization Machine to design a stackable architecture, empirically demonstrating that model performance improves favorably with an increased parameter count. Subsequent work has advanced this direction by adopting scalable architectures such as the MLP-Mixer~\citep{zhu2025rankmixer} or Transformer~\citep{gui2023hiformer,yan2025scaling,yu2025hhft}, achieving more effective and scalable designs. MTGR~\citep{han2025mtgr} and OneTrans~\citep{zhang2025onetrans} further advance this line by unifying sequence modeling and feature interaction within a single scalable Transformer, marking a key step toward unified modeling.

%% file: sections/preliminaries.tex
\section{Preliminaries} 
\label{sec:preliminary}

This section formalizes the CTR prediction task and describes the input processing pipeline that maps heterogeneous raw features into a unified token sequence and pre-trained content features.

\noindent \textbf{Problem Formulation.}
\Cref{fig: problem_setting} illustrates the CTR prediction task in recommendation systems.
Generally, in a single request, a CTR prediction model estimates the probability $\hat{y}_{u,c} \in [0, 1]$ that the user $u$ will click on the candidate item $c$ in the candidate set, usually formulated as a binary classification task. 
The input feature space of CTR prediction model comprises non-behavioral features $\mathcal{N}$, such as user profiles $\mathcal{U}$ and candidate features $\mathcal{C}$, as well as historical user behavior sequences $\mathcal{B}$. 
In practice, the user behavior sequences can be categorized into short-term user-specific behaviors $\mathcal{B}_{u}$ and lifelong user-specific behaviors $\mathcal{B}_l$. 
To efficiently handle lifelong behaviors, a General Search Unit (GSU)~\citep{wu2025muse} is employed to retrieve a candidate-specific subsequence $\mathcal{B}_c\subset \mathcal{B}_l$ containing items most relevant to the candidate $c$. 
Additionally, items in both the behaviors and the candidate set are structured records with
multiple modalities, including discrete ID embedding and content features (e.g., image and text).
Given these inputs, the CTR prediction model predicts the click score $\hat{y}_{u,c}$:
\begin{equation}
    \hat{y}_{u,c} = f(\mathcal{B}_{u}, \mathcal{B}_c, \mathcal{N}). \nonumber
\end{equation}
\noindent \textbf{Input Processing.}
In our model, all heterogeneous raw features are projected into a unified token sequence (details in~\Cref{sec: Features and Tokenization}). 
Behavioral sequences $(\mathcal{B}_{u}, \mathcal{B}_c)$ and non-behavioral features $\mathcal{N}$ are converted into token sequences $\mathbf{B}$ and $\mathbf{N}$, respectively. 
To incorporate content information, pre-trained content representations~\citep{sheng2024enhancing} for the candidate item and items in the behavior sequences are obtained via lookup, which remain frozen throughout training to preserve pre-trained semantics. 
The model input is formed by combining token sequences ($\mathbf{B}$, $\mathbf{N}$) with the frozen content features.

\section{Observations and Insights}
\label{sec: Observations and Insights} 
In this section, we investigate the fundamental distinctions between CTR prediction models and LLMs, revealing two key insights that guide the design of our architecture:

\begin{FindingBox}{Key Insights}
    \begin{itemize}[left=-3pt]
        \item 
        \textbf{Asymmetry in Information Density}\\ 
        Information density guides interaction priority. 
        \item \textbf{Modality-specific Priors}\\ Content signals are most effective as relational priors. 
    \end{itemize}
\end{FindingBox}

\subsection{Asymmetry in Information Density}
\label{sec: Information Density}

\begin{figure}[!t]
  \centering
  \begin{subfigure}[b]{0.22\textwidth}
    \includegraphics[width=\textwidth]{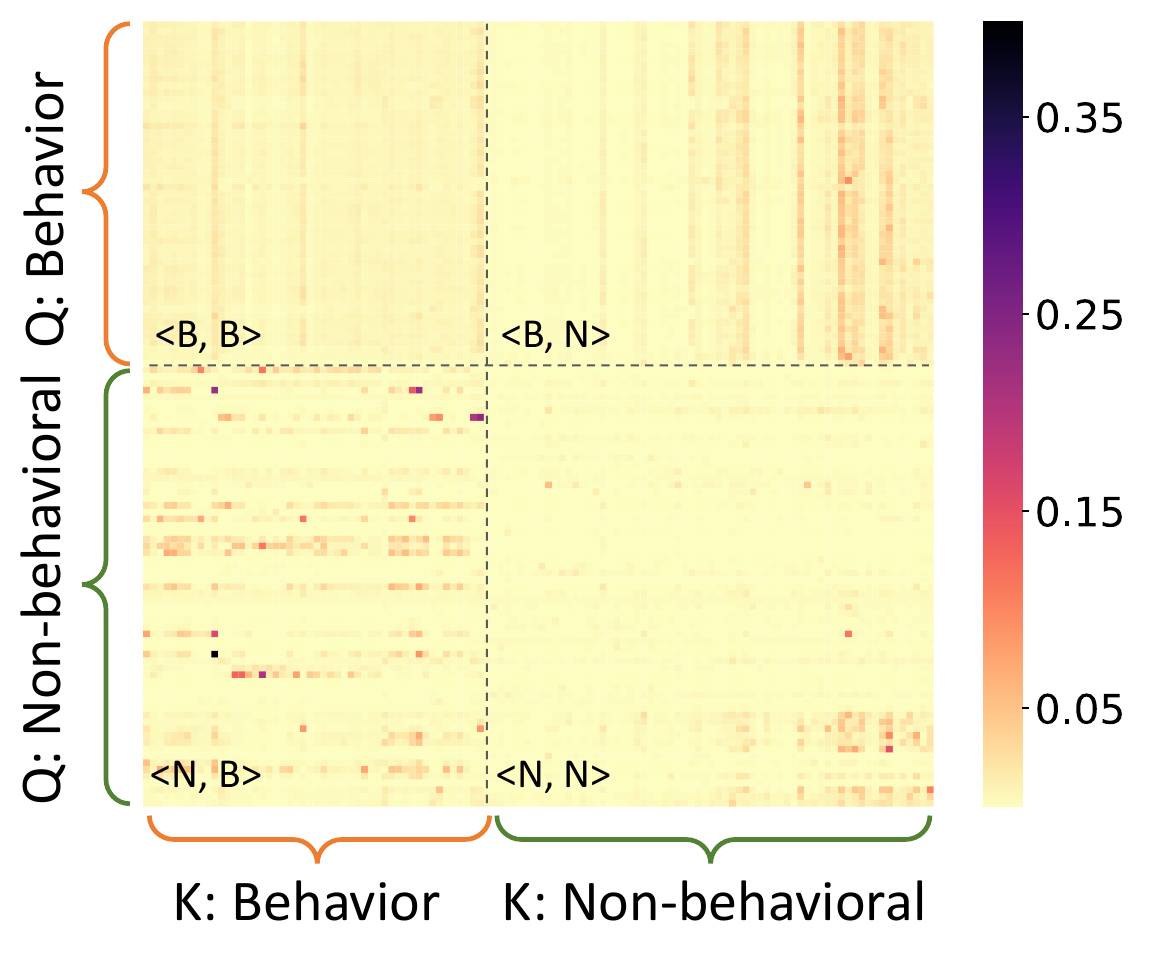}
    \vskip -0.05in
    \caption{Attention Matrix}
    \label{fig: attention_matrix}
  \end{subfigure}
  \begin{subfigure}[b]{0.22\textwidth}
    \includegraphics[width=\textwidth]{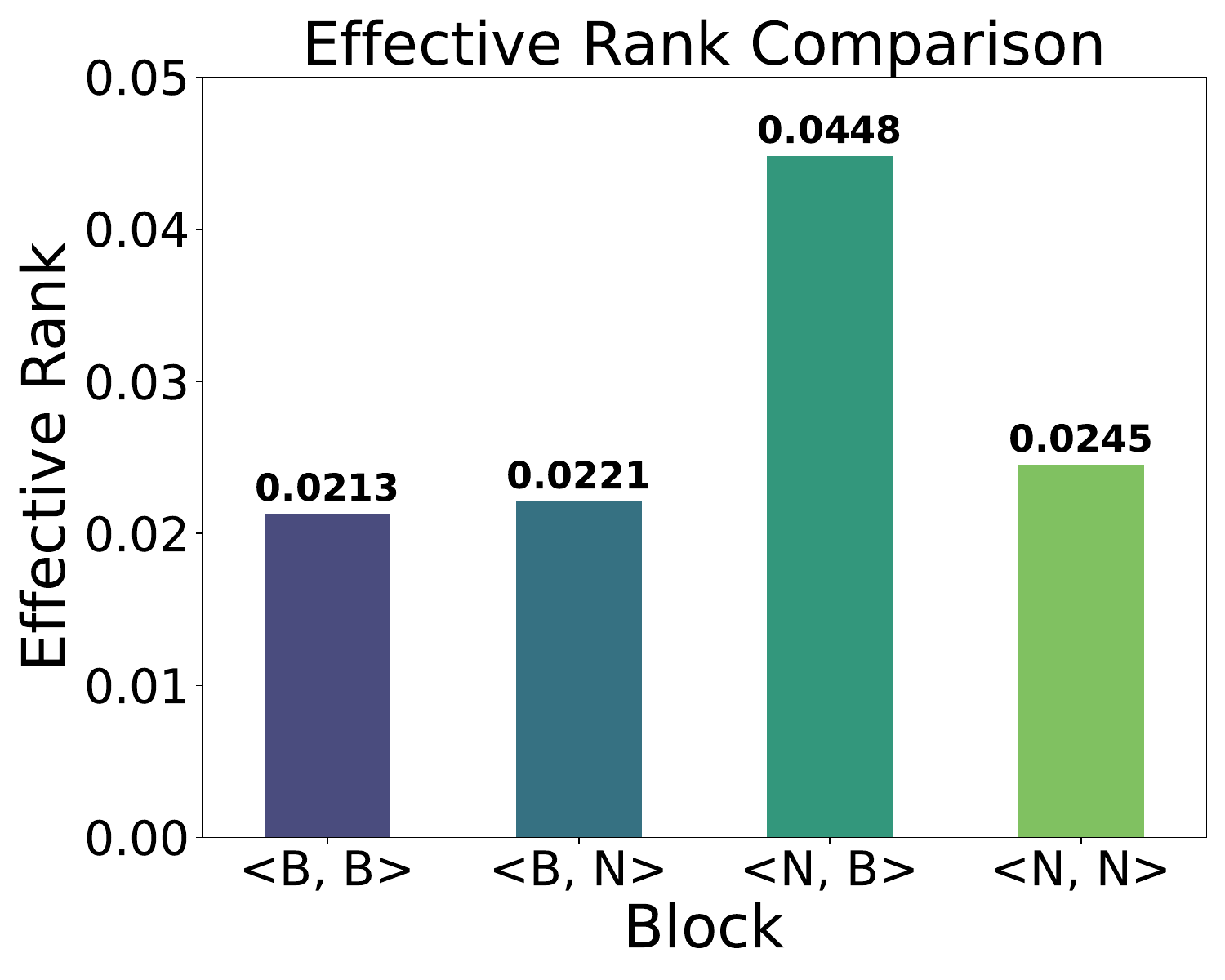}
    \vskip -0.02in
    \caption{Effective Rank}
    \label{fig: eranks}
  \end{subfigure}
  \vspace{-5pt}
  \caption{Visualization and effective rank of the attention matrix.
  For visual clarity, only the attention distributions of candidate-specific behavioral tokens and non-behavioral tokens are illustrated, as user-specific behaviors exhibit patterns similar to candidate-specific behaviors.}
  \vspace{-15pt}
\end{figure}

Unlike LLMs, which typically process homogeneous tokens, CTR prediction models operate on heterogeneous inputs that range from long behavioral sequences to compact, information-rich non-behavioral features. 
A critical question arises: do all token-to-token interactions contribute equally to the performance of the CTR prediction model? 
To examine this, we concatenate behavioral tokens $\mathbf{B}$ and non-behavioral tokens $\mathbf{N}$ into a single sequence $\mathbf{X} = [\mathbf{B}, \mathbf{N}]$ and employ self-attention to model their interactions: 
\begin{equation}
\begin{aligned}
    &(\mathbf{Q}, \mathbf{K}, \mathbf{V}) = (\mathbf{X}\mathbf{W}^{Q}, \mathbf{X}\mathbf{W}^{K}, \mathbf{X}\mathbf{W}^{V}),\\
    &\mathbf{X} \leftarrow \text{Softmax}(\frac{\mathbf{Q}\mathbf{K}^{\top}}{\sqrt{d}})\mathbf{V} + \mathbf{X}. \nonumber
\end{aligned}
\end{equation}
The attention matrix $\mathbf{A} = \text{Softmax}(\mathbf{QK}^\top / \sqrt{d})$ is visualized in \Cref{fig: attention_matrix}.
For clarity, let $\langle \mathbf{N}, \mathbf{B} \rangle$ denote the attention block where \emph{non-behavioral} tokens serve as \emph{queries} and \emph{behavioral} tokens serve as \emph{keys/values}, with other blocks defined analogously. 
We observe two properties. First, the attention matrix exhibits a distinct block-wise structure corresponding to different query-key pairs. Second, the $\langle \mathbf{N}, \mathbf{B} \rangle$ block displays more dispersed and diverse patterns, whereas other blocks suffer from rank collapse, where attention distributions become homogenized across different queries.

To quantify the differences in attention distributions across blocks, we adopt the effective rank~\citep{roy2007effective,rudelson2007sampling} as a proxy for information density. For any matrix $\mathbf{H}\in \mathbb{R}^{m \times n}$, the effective rank is defined as: 
$\text{erank}(\mathbf{H}) = \frac{1}{\max(m, n)}\|\mathbf{H}\|^{2}_{F}/\|\mathbf{H}\|^{2}_{2}$, where $\|\cdot\|_{F}$ and $\|\cdot\|_{2}$ denotes the Frobenius and spectral norm.
A higher effective rank indicates a more diverse information distribution and less redundancy. As in \Cref{fig: eranks}, the $\langle \mathbf{N}, \mathbf{B} \rangle$ block possesses a substantially higher erank compared to other blocks, implying that the interactions within this block are more information-intensive. 

We further validate these observations through a block-wise masking analysis. As demonstrated in \Cref{tab: masking}, masking the $\langle \mathbf{B}, \mathbf{B} \rangle$, $\langle \mathbf{B}, \mathbf{N} \rangle$, and $\langle \mathbf{N}, \mathbf{N} \rangle$ blocks results in negligible performance decay. In contrast, masking the $\langle \mathbf{N}, \mathbf{B} \rangle$ block leads to a significant drop. These observation confirms that interactions across heterogeneous CTR features are indeed unequal in their contributions, providing a strong motivation for the design of the Lightweight Cross-Attention (LCA) module (details in \Cref{sec: Lightweight Cross-Attention}), which prioritizes essential interactions while simplifying redundant computation. 

\subsection{Modality-Specific Priors} 
\label{sec: Information View} 
Content features provide rich semantic information that ID-based models
cannot capture. However, prior work shows that directly integrating
content features into ID-based models often yields suboptimal performance. 
A simpler yet more effective approach is to compute the cosine similarity between the content feature of the candidate item and those of items the user previously interacted with~\citep{yang2023courier,sheng2024enhancing}.
This strategy leverages pre-trained content features to capture fine-grained item relevance without the complexities of end-to-end multimodal training. 

To evaluate the effectiveness of different multimodal integration strategies, we compare three representative paradigms: (1) \textbf{Side Information}: Content features are concatenated with ID embeddings as additional attributes for each item. (2) \textbf{Semantic ID}~\citep{luo2025qarm}: Content features are discretized into cluster-based semantic IDs, which are then treated as traditional categorical features. (3) \textbf{SimTier}~\citep{sheng2024enhancing}: Content features are utilized to calculate similarity distributions between the target item and the user's historical interactions, serving as similarity-based signals.

As shown in \Cref{tab: mm_comparison}, the similarity-based SimTier approach significantly outperforms the direct integration methods, such as Side Information and Semantic ID. Interestingly, combining Side Information with SimTier leads to a slight performance decline compared to using SimTier alone, suggesting that different integration paradigms may be difficult to coordinate within a single optimization objective. This observation indicates that content features require specialized utilization mechanisms, moving away from naive token-level interaction. Guided by this insight, we propose the Content Sparse Attention (CSA) module (details in \Cref{sec: Content Sparse Attention}), which leverages content similarity to enable efficient and effective intra-sequence interaction. 

\begin{table}[!t]
  \centering
  \caption{Block-wise masking analysis of interactions between behavioral tokens and non-behavioral tokens.}
  \vspace{-3pt}
  \label{tab: masking}
  \begin{tabular}{lcc}
    \toprule
    \textbf{Masked Block}& $\Delta$\textbf{AUC} & $\Delta$\textbf{GAUC} \\
    \midrule
    Full Attention &-&- \\
    <$\mathbf{B}$, $\mathbf{B}$>& -0.01\% &-0.00\% \\
    <$\mathbf{B}$, $\mathbf{N}$>&  -0.03\% & -0.02\% \\
    <$\mathbf{N}$, $\mathbf{N}$> & -0.03\% & -0.02\% \\
    <$\mathbf{N}$, $\mathbf{B}$>&  \textbf{-0.20\%} & \textbf{-0.20\%}\\
    \bottomrule
  \end{tabular}
  \vspace{-10pt}
\end{table}

%% file: sections/method.tex

\section{Method}
Based on the insights derived in \Cref{sec: Observations and Insights}, we propose {\modelname}, an \acronym{E}fficiently \acronym{S}calable \acronym{T}ransformer-based architecture designed for industrial CTR prediction.
As illustrated in \Cref{fig: model_architecture}, {\modelname} organizes all raw features into a unified sequence without prior aggregation or merging (in \Cref{sec: Features and Tokenization}). 
The core of the {\modelname} block consists of two specialized modules: Lightweight Cross-Attention (LCA), which captures critical interactions between non-behavioral and behavioral inputs (in~\Cref{sec: Lightweight Cross-Attention}), and Content Sparse Attention (CSA), which integrates content information through intra-sequence content similarity (in~\Cref{sec: Content Sparse Attention}). 

\subsection{{\modelname} Overview}

\begin{table}[!t]
  \centering
  \caption{Comparison of multimodal integration approaches.}
  \label{tab: mm_comparison}
  \vspace{-5pt}
  \begin{tabular}{lcc}
    \toprule
    \textbf{Method}& \textbf{$\Delta$AUC} & \textbf{$\Delta$GAUC} \\
    \midrule
    ID-based Model& - & - \\
    Side Information& +0.44\% &+0.89\%  \\
    Semantic ID &+0.47\% &+0.64\% \\
    SimTier & +1.18\% & +1.58\%  \\
    SimTier+Side Information &+0.95\% &+1.47\% \\
    \bottomrule
  \end{tabular}
  \vspace{-10pt}
\end{table}

As mentioned in \Cref{sec:preliminary}, we denote the raw input features as $\mathcal{X} = (\mathcal{B}_{u}, \mathcal{B}_c, \mathcal{N})$. These heterogeneous inputs are projected into a unified space via specialized tokenizers or associated with pre-trained content features, yielding the non-behavioral token sequence $\mathbf{N}$, the behavioral token sequences $(\mathbf{B}_{u}, \mathbf{B}_c)$, and the corresponding frozen content features $(\mathbf{M}_{u}, \mathbf{M}_c)$, as elaborated in \Cref{sec: Features and Tokenization}. 
For notation brevity, let $\mathbf{B}$ denote the concatenation of the behavioral tokens $[\mathbf{B}_{u}, \mathbf{B}_c]$. As illustrated in \Cref{fig: model_architecture}, the {\modelname} block processes the behavioral token sequence $\mathbf{B}$ and the non-behavioral token sequence $\mathbf{N}$ via different modules. Non-behavioral tokens interact with behavioral tokens with the Lightweight Cross-Attention (LCA) module, while behavioral tokens integrate content features through the Content Sparse Attention (CSA) module. We utilize the superscript $s-1$ to represent the output of the $s$-th layer, where the computations are formulated as follows: 
\begin{equation}
\begin{aligned}
    &\mathbf{N}^{(s)} \leftarrow \text{LCA}\left(\text{Norm}(\mathbf{N}^{(s-1)}), \text{Norm}(\mathbf{B}^{(s-1)})\right) + \mathbf{N}^{(s-1)},\\
    &\mathbf{B}^{(s)} \leftarrow \text{CSA}\left(\text{Norm}(\mathbf{B}^{(s-1)})\right) + \mathbf{B}^{(s-1)}. \nonumber
\end{aligned}
\end{equation}

Following the attention mechanisms, the non-behavioral and behavioral tokens are further processed by Feed-Forward Networks (FFNs~\citep{DBLP:journals/corr/abs-2002-05202}), respectively: 
\begin{equation}
\begin{aligned}
    &\mathbf{N}^{(s)} \leftarrow \text{PerToken-FFN}\left(\text{Norm}(\mathbf{N}^{(s)})\right) + \mathbf{N}^{(s)},\\
    &\mathbf{B}^{(s)} \leftarrow \text{Shared-FFN}\left(\text{Norm}(\mathbf{B}^{(s)})\right) + \mathbf{B}^{(s)}, \nonumber
\end{aligned}
\end{equation}
where \text{PerToken-FFN} denotes independent networks tailored to each non-behavioral token, and \text{Shared-FFN} represents a universal network applied uniformly across all behavioral tokens. 

After the $S$ layers of stacked {\modelname} blocks, the behavioral tokens $\mathbf{B}^{(S)}$ are mean-pooled and concatenated with the non-behavioral tokens $\mathbf{N}^{(S)}$ to form the final input representation $\mathbf{X}^{(S)}$. We utilize a Multilayer Perceptron (MLP) to predict the probability $\hat{y}_{u, c}$ that the user $u$ will click on the candidate item $c$: 
\begin{equation}
\begin{aligned}
    &\mathbf{X}^{(S)} = [\mathbf{N}^{(S)}, \text{Mean}(\mathbf{B}^{(S)})], \\
    &\hat{y}_{u, c} = \text{Sigmoid}\left(\text{MLP}(\mathbf{X}^{(S)})\right). \nonumber
\end{aligned}
\end{equation}
Given label $y_{u, c}$, the model is trained with the cross-entropy loss: 
\begin{equation}
\begin{aligned}
    &\mathcal{L} = -\sum_{u, c} \left[y_{u, c}\log\hat{y}_{u,c} + (1-y_{u,c})\log(1-\hat{y}_{u,c})\right]. \nonumber
\end{aligned}
\end{equation}

\subsection{Tokenization}
\label{sec: Features and Tokenization}
To enable fully unified modeling, we project all heterogeneous raw inputs into tokens of a unified dimension $d$.
This process ensures that diverse signals, ranging from categorical IDs to user history, are aligned into a unified embedding space while preserving their fine-grained details.
Furthermore, frozen content features are incorporated to supplement the model with rich semantics.

\subsubsection{Non-behavioral tokenization.}  
For each raw non-behavioral feature $n_i \in \mathcal{N}$, we employ a feature-specific tokenizer to map it into a token $\mathbf{n}_i \in \mathbb{R}^d$ without aggregation or merging.
All resulting non-behavioral tokens are concatenated into the sequence $\mathbf{N}$: 
\begin{equation}
\begin{aligned}
    &\mathbf{n}_{i} = \text{MLP}_{i}(n_i), 
    n_{i}\in \mathcal{N}, \\
    &\mathbf{N} = [\mathbf{n}_1, \ldots, \mathbf{n}_{L_\mathbf{N}}], \mathbf{N}\in\mathbb{R}^{L_{\mathbf{N}}\times d}, \nonumber
\end{aligned}
\end{equation}
where $L_N$ denotes the number of non-behavioral features. 
This design enables a plug-and-play feature expansion via warm-starting, allowing new features to be added without architectural changes or parameter reinitialization.

\subsubsection{Behavioral tokenization.} 
For notational simplicity, we use $\mathcal{B}_{\gamma}$ ($\gamma \in \{s,c\}$) to denote either
the user-specific behavior sequence $\mathcal{B}_{u}$ or the candidate-specific behavior sequence $\mathcal{B}_{c}$.
For each sequence $\mathcal{B}_{\gamma}$, we adopt a sequence-specific tokenizer to map the ID embedding of the item $b_{\gamma, i}$ within the sequence into a token $\mathbf{b}_{\gamma, i}$ with the dimension $d$.
These tokens are concatenated into the corresponding behavioral token sequence $\mathbf{B}_{\gamma}$:
\begin{equation}
\begin{aligned}
    &\mathbf{B}_{\gamma} = [\text{MLP}_{\gamma}(b_{\gamma, 1}), \ldots, \text{MLP}_{\gamma}(b_{\gamma, L_{\mathbf{B}_{\gamma}}})], \mathbf{B}_{\gamma}\in\mathbb{R}^{L_{\mathbf{B}_{\gamma}}\times d}. \nonumber
\end{aligned}
\end{equation}
We use $\mathbf{B}=[\mathbf{B}_{u}, \mathbf{B}_{c}] \in \mathbb{R}^{L_{\mathbf{B}}\times d}$ to denote the concatenation of all behavioral token sequences, where $L_{\mathbf{B}} = L_{\mathbf{B}_{u}}+L_{\mathbf{B}_{c}}$. 

\subsubsection{Content feature mapping.} 
To incorporate content features, we map each item in the behavioral sequences $\mathcal{B}_{u}$ and $\mathcal{B}_c$ to its corresponding pre-trained content feature $\mathbf{m} \in \mathbb{R}^{d_M}$ through a lookup operation. This results in the content sequences $\mathbf{M}_{u} \in \mathbb{R}^{L_{\mathbf{B}_{u}} \times d_M}$ and $\mathbf{M}_c \in \mathbb{R}^{L_{\mathbf{B}_{c}} \times d_M} $, which provide a semantic perspective that complements the ID-based tokens.
These features, together with the candidate content feature, are kept frozen during training. Freezing preserves the semantics of the pretrained feature and reduces optimization instability that can arise when jointly learning high-dimensional dense vectors and discrete ID embeddings.

\subsection{Lightweight Cross-Attention}
\label{sec: Lightweight Cross-Attention}
Motivated by the observation in \Cref{sec: Observations and Insights} that interactions between non-behavioral and behavioral tokens are the most informative, we propose Lightweight Cross-Attention (LCA). This module is designed to prioritize the critical interactions while discarding other redundant interactions. 
In LCA, non-behavioral tokens $\mathbf{N} \in \mathbb{R}^{L_N \times d}$ serve as queries, while the concatenated behavioral tokens $\mathbf{B} = [\mathbf{B}_{u}, \mathbf{B}_c] \in \mathbb{R}^{L_B \times d}$ serve as the keys and values: 

\begin{equation}
\begin{aligned}
    &(\mathbf{Q}_{\mathbf{N}}, \mathbf{K}_{\mathbf{B}}, \mathbf{V}_{\mathbf{B}}) = \left(\phi(\mathbf{N},\mathbf{W}^{Q}), \mathbf{B}\mathbf{W}^{K}, \mathbf{B}\mathbf{W}^{V}\right),\\
    &\text{LCA}(\mathbf{N}, \mathbf{B}) = \phi\left(\text{Softmax}(\frac{\mathbf{Q}_{\mathbf{N}}\mathbf{K}_{\mathbf{B}}^{\top}}{\sqrt{d}})\mathbf{V}_{\mathbf{B}}, \mathbf{W}^{O}\right), \nonumber
\end{aligned}
\end{equation}
where $\phi(\cdot)$ denotes token-wise matrix multiplication. Specifically, $\mathbf{W}_Q, \mathbf{W}_O \in \mathbb{R}^{L_N \times d \times d}$ are token-specific projection weights, where each non-behavioral token maintains its own independent set of parameters to capture unique feature signals. Meanwhile, $\mathbf{W}_K, \mathbf{W}_V \in \mathbb{R}^{d \times d}$ are shared projection weights applied uniformly across all behavioral tokens to ensure computational scalability.

LCA offers significant advantages for industrial deployment. Since the user-specific behavioral sequence $\mathbf{B}_{u}$ is agnostic of candidate items, its corresponding projections $\mathbf{B}_{u}\mathbf{W}^K$ and $\mathbf{B}_{u}\mathbf{W}^V$ can be computed once and reused across all candidate items within a single ranking request. This user-candidate decoupled computation significantly reduces the inference latency. 

\noindent \textbf{Computational Complexity.} 
The complexity of LCA arises from projections and attention. Compared with the standard self-attention, LCA minimizes computational redundancy by simplifying interactions and supporting {user-candidate} decoupled computation. Formally, the complexity is reduced as follows: 
\begin{align}
    \text{Projection:} \quad &O\left(3(L_N + L_B)d^2\right) \rightarrow O(L_N d^2 + 2L_B d^2), \nonumber\\
    \text{Attention:} \quad &O\left((L_N + L_B)^2 d\right) \rightarrow O(L_N L_B d).\nonumber
\end{align} 
In a typical industrial setting where $L_B \gg L_N$, this reduction from quadratic to bilinear complexity enables EST to handle much longer behavioral sequences without exceeding latency budgets.

{
\begin{algorithm}[t]
\caption{Content Sparse Attention (CSA)}
\label{alg: CSA}
\KwIn{Behavior tokens $\mathbf{B}_{\gamma} \in \mathbb{R}^{L_{\mathbf{B}_{\gamma}} \times d}$, frozen content features $\mathbf{M}_{\gamma} \in \mathbb{R}^{L_{\mathbf{B}_{\gamma}} \times d_M}$, top-K hyperparameter $K$}
\KwOut{CSA output $\mathbf{O}_{\mathbf{B}_{\gamma}} \in \mathbb{R}^{L_{\mathbf{B}_{\gamma}} \times d}$}

\tcp{Compute content similarity matrix}
$\mathbf{G}_{\gamma} \gets \mathbf{M}_{\gamma}\mathbf{M}_{\gamma}^{\top}$ 
\tcp*{${L_{\mathbf{B}_{\gamma}} \times L_{\mathbf{B}_{\gamma}}}$}

\tcp{Row-wise top-$K$ indices and similarities}
$\mathbf{\Omega}_{\gamma}, \mathbf{E}_{\gamma} \gets \text{TopK}(\mathbf{G}_{\gamma}, K)$ 
\tcp*{${L_{\mathbf{B}_{\gamma}} \times K}$}

\tcp{Gather top-$K$ items based on indices}
$\mathbf{B}_{\gamma}^{\text{gth}} \gets \text{Gather}(\mathbf{B}_{\gamma}, \mathbf{\Omega}_{\gamma})$ 
\tcp*{${L_{\mathbf{B}_{\gamma}} \times K \times d}$}

\tcp{Sparse attention with content similarity}
$\mathbf{E}_{\gamma} \gets \text{Unsqueeze}(\mathbf{E}_{\gamma}, -1)$ 
\tcp*{${L_{\mathbf{B}_{\gamma}} \times K \times 1}$}
$\mathbf{O}_{\mathbf{B}_{\gamma}} \gets \text{Sum}(\mathbf{B}_{\gamma}^{\text{gth}} \odot \mathbf{E}_{\gamma}, \text{axis}=1)$ 
\tcp*{${L_{\mathbf{B}_{\gamma}} \times d}$}

\Return $\mathbf{O}_{\mathbf{B}_{\gamma}}$
\end{algorithm}
}
\subsection{Content Sparse Attention}
\label{sec: Content Sparse Attention}
Guided by the insight that content similarity effectively enriches ID-based models, we propose Content Sparse Attention (CSA). 
Unlike conventional self-attention that relies on learnable parameters to capture token relationships, CSA leverages the fixed semantic structures of pre-trained content features to guide intra-sequence interactions. This design yields a training-free, computationally efficient module, with full details provided in~\Cref{alg: CSA}.
To maintain the user-candidate decoupled computation, CSA is performed independently on the user-specific sequence $\mathcal{B}_{u}$ and the candidate-specific sequence $\mathcal{B}_c$. For brevity, we denote either sequence as $\mathcal{B}_\gamma$, with $\gamma \in \{s, c\}$.

Given the behavior sequence $\mathcal{B}_{\gamma}$, 
we utilizes the corresponding normalized content representations $\mathbf{M}_{\gamma} \in \mathbb{R}^{L_{\mathbf{B}_{\gamma}\times d_{\mathbf{M}}}}$ to compute the intra-sequence  similarity matrix $\mathbf{G}_{\gamma}= \mathbf{M}_{\gamma}\mathbf{M}_{\gamma}^{\top}$, where each element $\mathbf{G}_{\gamma}[i, j]$ represents the content similarity between the $i$-th item and $j$-th item within the behavior sequence $\mathcal{B}_{\gamma}$.
Derived from frozen content features, the similarity matrix $\mathbf{G}_{\gamma}$ does not involve backpropagation and can be reused by all layers.
Carrying the semantic relationships, $\mathbf{G}_{\gamma}$ can serve as a training-free attention matrix to model interactions within the behavior:
\begin{equation}
\label{eq: CSA}
\begin{aligned}
    &\mathbf{O}_{\mathbf{B}_{\gamma}} = \mathbf{G}_{\gamma}\mathbf{B}_{\gamma}, \mathbf{O}_{\mathbf{B}_{\gamma}}\in \mathbb{R}^{L_{\mathbf{B}_{\gamma}}\times d}.
\end{aligned}
\end{equation}
Through \Cref{eq: CSA},  the ID-based behavioral token of items attend to each other guided by their content similarities.
This achieves a complementarity between co-occurrence relationships captured by ID features and semantic relationships inherent in content features.

Although simple, \Cref{eq: CSA} is computationally expensive due to its quadratic complexity of $O(L_{\mathbf{B}_{\gamma}}^2 d)$. 
To reduce this cost, we apply row-wise top-$K$ sparsification to the similarity matrix $\mathbf{G}_{\gamma}$.
By restricting each item to attend only to the $K$ most similar items, we implement Content Sparse Attention.
Consequently, the complexity is reduced to $O(L_{\mathbf{B}_{\gamma}}Kd)$,
scaling linearly with $L_{\mathbf{B}_{\gamma}}$.

\noindent\textbf{Computational complexity.}
The computational complexity arises from the computation of the content attention matrix $\mathbf{G}_{\gamma}$ and the sparse attention in \Cref{alg: CSA}:
\begin{equation}
\left\{
\begin{aligned}
  & \text{Content attention Matrix}: &&O(L_{\mathbf{B}_{\gamma}}d_{M}^2), \nonumber\\
  & \text{Sparse Attention}: &&O(L_{\mathbf{B}_{\gamma}}Kd). \nonumber
\end{aligned}
\right.
\end{equation}
The design of CSA yields the following benefits:
(1) \textbf{Parallel precomputation}: Both the construction of the content attention matrix and top-$K$ sparsification can be performed in parallel with other feature processing operations in the ranking stage, and reused by each EST layer, incurring negligible latency.
(2) \textbf{User-candidate decoupled computation}: 
CSA processes  $\mathbf{B}_{u}$ and $\mathbf{B}_{c}$ independently.
This allows user-specific behavior computations to be shared across multiple candidate items.
(3) \textbf{Linear complexity}: In all our experiments, we fix $K=5$, which is much smaller than the sequence length $L_{\mathbf{B}_{\gamma}}$. With top-$K$ sparsification, the computational complexity therefore scales linearly with the sequence length $L_{\mathbf{B}_{\gamma}}$.

\subsection{Training and Deployment}
\label{sec: Training and Deployment}
Despite strong gains through architectural redesign, it remains difficult to train and deploy \modelname~from scratch in industry. The online production model has been trained for years on \emph{several trillion samples}, making its performance hard to exceed. We therefore initialize sparse parameters from the historical production model and train only dense parameters from scratch. After training on hundreds of billions of samples, EST surpasses the production model by \textbf{$0.74\%$} in GAUC and sustains improvement with additional data.
To maximize data utility, we adopt a \textbf{multi-epoch training strategy}. However, recommendation systems commonly suffer performance degradation at the start of the second epoch due to overfitting, named one-epoch limitation~\citep{zhang2022towards}. Inspired by~\citet{liu2023multi} and \citet{fan2024multi}, we implement an asynchronous paradigm:
\begin{alignat}{2}
    \text{Sparse Parameters}: \boldsymbol{\theta}_{\mathrm{s}}^{(e)} &\xleftarrow{\text{init}} \boldsymbol{\theta}_{\mathrm{s}}^{(0)}, 
    &\quad& \forall\, e \geq 1, \nonumber\\[6pt]
    \text{Dense Parameters}:\boldsymbol{\theta}_{\mathrm{d}}^{(e)} &\xleftarrow{\text{init}}
    \left\{
    \begin{array}{@{}l@{}}
        \text{random}, \\
        \boldsymbol{\theta}_{\mathrm{d}}^{(e-1)},
    \end{array}
    \right.
    &\quad&
    \begin{array}{@{}l@{}}
        e = 1, \\
        e \geq 2,
    \end{array}
    \nonumber
\end{alignat}
where ($\boldsymbol{\theta}_{\mathrm{s}}^{(e)}$, $\boldsymbol{\theta}_{\mathrm{d}}^{(e)}$) denotes the sparse and dense parameters at the $e$-th epoch, while the sparse initialization  ${\boldsymbol{\theta}}_{\mathrm{s}}^{(0)}$ is from the historical production model. In each epoch, we reset sparse parameters to $\boldsymbol{\theta}_{ \mathrm{s}}^{(0)}$ while carrying dense parameters from the previous epoch, with random dense initialization only in epoch one. This design mitigates overfitting and brings an additional \textbf{$0.44\%$} GAUC gain.

%% file: sections/experiment.tex

\section{Experiments}

\begin{table*}[!t]
  \caption{CTR Prediction Performance. {Best} and {second-best} result are highlighted in {bold} and {underline}.}
  \label{tab: main_exp}
  \conferenceonly{\vspace{-10pt}}
  \label{tab:freq}
  \setlength{\tabcolsep}{9pt}
  \begin{tabular}{l|l|cccccc}
    \toprule
    \textbf{Modeling Paradigm} & \textbf{Method} &\textbf{AUC} &\textbf{$\Delta$AUC} &\textbf{GAUC}  &\textbf{$\Delta$GAUC} &\textbf{Params(M)} &\textbf{GFLOPs} \\
    \midrule
    \textbf{Base Model} & MLP& 0.6895 &- & 0.6459 &- & 83.13 & 9.64 \\
    \midrule
    \multirow{2}{*}{\textbf{Traditional DNN Model}} 
    &AutoInt& 0.6907 &0.18\% & 0.6477 &0.28\% & 5.20 &1.10 \\
    &DCNv2 & 0.6916 &0.31\%  & 0.6482 &0.35\% & 15.18 & 1.49  \\
    
    \midrule
    \multirow{2}{*}{\textbf{Scalable Hierarchical Modeling}} 
    &RankMixer & 0.6914 &0.28\% & 0.6482 &0.35\% & 136.84 & 15.46 \\
    &HiFormer & 0.6928 &0.49\% & 0.6493 &0.53\%& 113.78 & 14.17 \\
    \midrule
    \multirow{5}{*}{\textbf{Scalable Unified Modeling}} 
     & MTGR & 0.6917 &0.32\% & 0.6474 &0.23\% & 118.79 & 10.57 \\
     & OneTrans & 0.6921 &0.38\% & 0.6477 &0.27\% & 111.98 & 15.94  \\
     & OneTrans-D & 0.6941 &0.68\% & {0.6500} &0.64\% & 112.06 & 16.19\\
     & OneTrans-F & \underline{0.6955} &\underline{0.87\%} & \underline{0.6507} &\underline{0.75\%} & 110.54 & 52.79\\
     & \textbf{{\modelname}} & \textbf{0.6963} &\textbf{0.99\%} & \textbf{0.6515} &\textbf{0.87\%} & 103.81 & 18.48 \\
  \bottomrule
\end{tabular}
    \conferenceonly{\vspace{-10pt}}
\end{table*}

Through offline evaluations and online tests, we aim to answer the following Research Questions (RQs):
\begin{itemize}[left=0pt]
\item {\textbf{RQ1}}: \textbf{Performance Comparison}. Does the proposed {\modelname} outperform state-of-the-art CTR prediction models under comparable computation overhead and model capacity?
\item {\textbf{RQ2}}: \textbf{Ablation Study}. How do individual components contribute to overall performance and efficiency?
\item {\textbf{RQ3}}: \textbf{Scaling Law}. As we scale computation overhead and model capacity,
does {\modelname} exhibit a well-behaved scaling law?
\item {\textbf{RQ4}}: \textbf{Online A/B Testing}. How much real-world improvement does EST deliver when deployed in a production system?
\end{itemize}

\subsection{Experimental Setup}

\subsubsection{Datasets}
For offline evaluation, we use a CTR dataset sampled from Taobao production logs, comprising over 1 billion user impressions collected over six days.
To assess scalability, we further extend our experiments to a larger dataset with a richer feature set, spanning over 3.5 billion user impressions collected over the same period.
Our offline datasets contain hundreds of heterogeneous features, including non-behavioral features and user behavior sequences.
Specifically, the short-term sequence $\mathcal{B}_{u}$ and lifelong behavior sequence $\mathcal{B}_{l}$ are truncated to lengths of 300 and 5,000, respectively.
For the lifelong behavior sequence, the GSU module \cite{wu2025muse} extracts the top-50 candidate-specific sequence $\mathcal{B}_{c}$.

\subsubsection{Evaluation Metrics}
To evaluate model performance, we adopt AUC (Area Under the ROC Curve) and GAUC (Group-level AUC) as effectiveness metrics, and Params and Inference GFLOPs as efficiency metrics.
\textbf{AUC}: A higher AUC indicates stronger overall ranking capability.  
\textbf{GAUC}: GAUC measures the quality of intra-user ranking and has been shown to correlate more consistently with online performance~\cite{sheng2021one,sheng2023joint,zhou2018deep}. An improvement of 0.001 in GAUC is generally considered a confidently significant gain. We report relative improvements as $\Delta$AUC and $\Delta$GAUC. 
\textbf{Params}: The number of parameters in the dense part of the model, excluding sparse embedding tables.  
\textbf{Inference GFLOPs}: Computational costs measured in giga floating-point operations (GFLOPs) for processing a single batch of 60 samples during inference.

\subsubsection{Baselines}
We compare against representative state-of-the-art methods, categorized by modeling paradigm:
\begin{itemize}[left=0pt]
    \item  \textbf{Base model}: a vanilla \emph{MLP}-based feature interaction model.
    \item  \textbf{Traditional DNN Model}: \emph{AutoInt} \cite{song2019autoint} is a representative attention-based feature interaction model. \emph{DCNv2} \cite{wang2021dcn} captures high-order feature interactions via gated fusion. 
    \item  \textbf{Scalable Hierarchical Modeling}:  \emph{RankMixer} \cite{zhu2025rankmixer} introduces an effective architecture via token-mixer. \emph{HiFormer} \cite{gui2023hiformer} applies heterogeneous attention over aggregated inputs.
    \item \textbf{Scalable Unified Modeling}: \emph{MTGR} \cite{han2025mtgr} stacks HSTU \cite{zhai2024actions} blocks for unified sequence modeling. \emph{OneTrans} \cite{zhang2025onetrans} is a transformer-based framework with a pyramid stack.
\end{itemize}
Note that MTGR and OneTrans simply aggregate the candidate-specific behavior sequence into one token via pooling. 
To explore the impact of the candidate-specific behavior, we implement two enhanced variants: 
\textbf{OneTrans-D}, which uses \underline{D}IN to aggregate the candidate-specific sequence, 
and \textbf{OneTrans-F}, which directly concatenates the \underline{F}ull candidate-specific behavior sequence.

\subsubsection{Implementation Details}
All models are implemented with RecIS~\cite{ali2025recis} and optimized with the AdamW optimizer. 
All baselines use the same learning rate of 0.0004.
The offline experiments are conducted on a distributed training system with 32 PPUs, which is an internal GPU architecture developed by Alibaba.
We use 128 PPUs for the scaling experiments.
The batch size per PPU is 1000.
For a fair comparison, EST and other scalable baselines share the same architecture: 6 layers, a token dimension of 128, and approximately 100M parameters.
All baselines process the output of the last layer similarly to EST by mean-pooling the behavioral tokens and concatenating them with the non-behavioral tokens.

For the Base model, Traditional DNN models, and Scalable Hierarchical Modeling paradigm, we adopt DIN \cite{zhou2018deep} modules to compress behavior sequences. 
Regarding OneTrans-D, we adopt the DIN module to compress the candidate-specific sequence before being concatenated with the user behavior sequence and non-behavioral tokens. 
OneTrans-F directly concatenates all the behavioral and non-behavioral tokens into a single token sequence.

All baselines adopt SimTier~\citep{sheng2024enhancing} to convert the degree of content similarity between the candidate item and the behavior sequences into a specific type of non-behavioral token included in the non-behavioral token sequence $\mathbf{N}$. 
Building upon this, {\modelname} employs CSA to exploit intra-sequence content similarity, further enhancing the utility of content features.

\subsection{Performance Comparison (RQ1)}
\label{sec: performance}
\Cref{tab: main_exp} shows the performance of all models on the offline evaluation dataset.
{\modelname} outperforms all the baselines under comparable computational costs.
By comparing model performance, we derive three insightful conclusions.
\textbf{First, token-specific parameterized attention modules excel at capturing interactions among heterogeneous features.}
Comparing the Traditional DNN Model and the Scalable Hierarchical Modeling, HiFormer achieves the best performance gains by leveraging the heterogeneous attention layers. This observation underscores the potential of token-specific parameterized attention to model the intricate interactions among heterogeneous features.

 \textbf{Second, effective modeling of candidate-specific behavior sequences is critical for performance.}
Although OneTrans and MTGR enable thorough feature interactions through unified modeling, they fall short of HiFormer due to inadequate candidate-specific behavior modeling. 
However, by integrating a DIN module for candidate-specific behavior sequence compression, OneTrans-D notably outperforms HiFormer. 
Furthermore, by introducing the full candidate-specific behavior sequence, OneTrans-F obtains further improvements, but at the cost of a significant increase in computational overhead.
These results highlight that fully leveraging the candidate-specific behavior sequence is essential,
and a careful trade-off with computational costs is necessary.

\textbf{Third, {\modelname} improves performance without sacrificing efficiency.}
{\modelname} surpasses OneTrans by $0.59\%$ in relative GAUC while maintaining similar computational cost. Compared with OneTrans-F, it achieves a $0.12\%$ GAUC gain with $65\%$ less computation. This efficiency comes from reducing redundant interactions, which preserves effective modeling of candidate-specific behaviors and supports content fusion.

\begin{figure}[!t]
    \centering
    \vspace{-15pt}
    \includegraphics[width=0.7\linewidth]{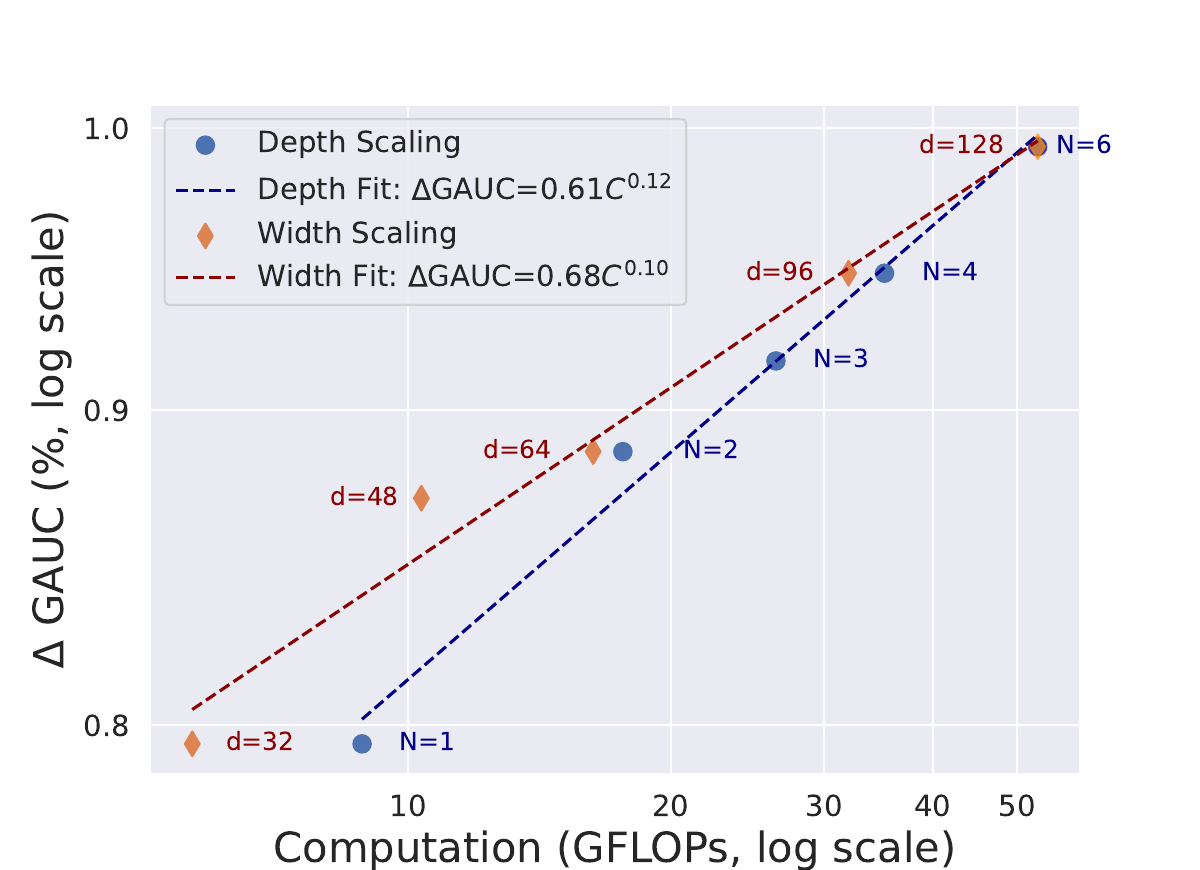}
    \caption{The power-law relationship between computation overhead and GAUC. As the model depth increases, the best-fit curve: $\Delta\text{GAUC} = 0.61 \times C^{0.12}$. As the model width increases, the best-fit curve: $\Delta\text{GAUC} = 0.68 \times C^{0.10}$.}
    \label{fig: scaling_flops}
\end{figure}

\begin{table}[h]
  \centering
  \caption{Ablation results.}
  \label{tab: ablation_study}
  \conferenceonly{\vspace{-10pt}}
  \setlength{\tabcolsep}{3pt}
  \begin{tabular}{l|ccc|ccc}
    \toprule
    \textbf{Method} & \textbf{FA} & \textbf{LCA} & \textbf{CSA} & \textbf{$\Delta$AUC} & \textbf{$\Delta$GAUC} & \textbf{$\Delta$GFLOPs} \\
    \midrule
    Full-Attention & \checkmark &- &- & 0.6951 & 0.6506  & 66.27 \\
    \midrule
    \multirow{2}{*}{{\modelname}}  & - & \checkmark &- &-0.01\%&-0.03\%	&-72.23\%\\
    &- & \checkmark & \checkmark&+0.17\% &+0.14\%&-72.11\% \\
    \bottomrule
  \end{tabular}
\end{table}

\subsection{Ablation Study (RQ2)}
We anchor our comparison on Full-Attention, which adopts full self-attention (FA) to implement unified modeling. 
Apart from this, data arrangement, the architectures of the FFN module, and hyperparameter settings remain consistent with EST.
The ablations in \Cref{tab: ablation_study} show that:
(1) Compared to full self-attention, our LCA reduces computation by 72.23\% with only a 0.03\% drop in performance. 
This further validates our conclusion in \Cref{sec:preliminary}: the full self-attention contains redundant interactions, while the interactions between non-behavioral features and behavioral sequences are the most informative. These results demonstrate that the design of LCA is both efficient and effective.
(2) Building upon the LCA module, CSA further boosts performance by 0.14\%, demonstrating that leveraging content similarity within behavior sequences provides additional gains. 
Meanwhile, through the sparsification design, CSA ensures efficiency with only a marginal increase in computation.
In summary, the synergy between LCA and CSA yields reliable and efficient improvements.

\subsubsection{Sensitivity Analysis}
\label{sec: Sensitivity Analysis}
In this section, we investigate the impact of the sparsification hyperparameter $K$ in CSA. We conduct comparative experiments under two settings: (1) disabling the top-$K$ sparsification of CSA, \emph{i.e.}, non-sparse CSA, and (2) varying $K \in \{5, 20, 60, 100\}$.
From the results in \Cref{tab:sensitivity_K}, we draw the following two conclusions:
(1) \textbf{Sparsification improves both effectiveness and efficiency.}
By comparing CSA with its non-sparse variant, we observe that top-$K$ sparsification is not merely a compromise adopted for computational efficiency. On the contrary, it simultaneously improves model performance while reducing computational overhead.
(2) \textbf{CSA is robust across different $K$.}
The performance of CSA remains stable across a wide range of $K$ values, demonstrating its robustness. To minimize computational overhead, we fix $K\!=\!5$ throughout our experiments.

\begin{table}[t]
  \centering
  \caption{Sensitivity analysis of the hyperparameter $K$.}
  \label{tab:sensitivity_K}
  \begin{tabular}{lcc}
    \toprule
    Method & AUC & GAUC \\
    \midrule
    EST w/ non-sparse CSA & 0.6951 & 0.6511 \\
    \midrule
    EST w/ Top-5 CSA   ($K\!=\!5$)   & \textbf{0.6963} & 0.6515 \\
    EST w/ Top-20 CSA  ($K\!=\!20$)  & 0.6960 & \textbf{0.6517} \\
    EST w/ Top-60 CSA  ($K\!=\!60$)  & 0.6956 & 0.6515 \\
    EST w/ Top-100 CSA ($K\!=\!100$) & 0.6955 & 0.6515 \\
    \bottomrule
  \end{tabular}
\end{table}

\subsubsection{User Activity-Level Analysis}
\label{sec: Stratified Analysis}
While \Cref{tab: main_exp} demonstrates EST's overall effectiveness, a fine-grained stratified analysis is needed to examine its robustness across different user activity levels.
As shown in \Cref{tab:activity_level}, EST consistently outperforms OneTrans-F across all activity levels. Notably, the most substantial improvement of $0.1569\%$ is observed on low-activity users (behavior length $< 500$). This validates that the semantic relations introduced by CSA are especially beneficial when limited history weakens ID embeddings.

\begin{table}[t]
  \centering
  \caption{User Activity-Level Analysis. $\Delta$GAUC represents the relative GAUC improvement of EST.}
  \label{tab:activity_level}
  \begin{tabular}{lccc}
    \toprule
    Behavior Length & OneTrans-F & EST & $\Delta$GAUC \\
    \midrule
    $[0, 500)$      & 0.6446 & 0.6456 & \textbf{+0.1569\%} \\
    $[500, 1000)$   & 0.6493 & 0.6501 & +0.1333\% \\
    $[1000, 2000)$  & 0.6491 & 0.6499 & +0.1278\% \\
    $[2000, 3000)$  & 0.6493 & 0.6501 & +0.1188\% \\
    $[3000, 4000)$  & 0.6491 & 0.6496 & +0.0754\% \\
    $[4000, 5000]$  & 0.6523 & 0.6530 & +0.1027\% \\
    \bottomrule
  \end{tabular}
\end{table}

\begin{figure}[!t]
    \centering
    \vspace{-15pt}
    \includegraphics[width=0.7\linewidth]{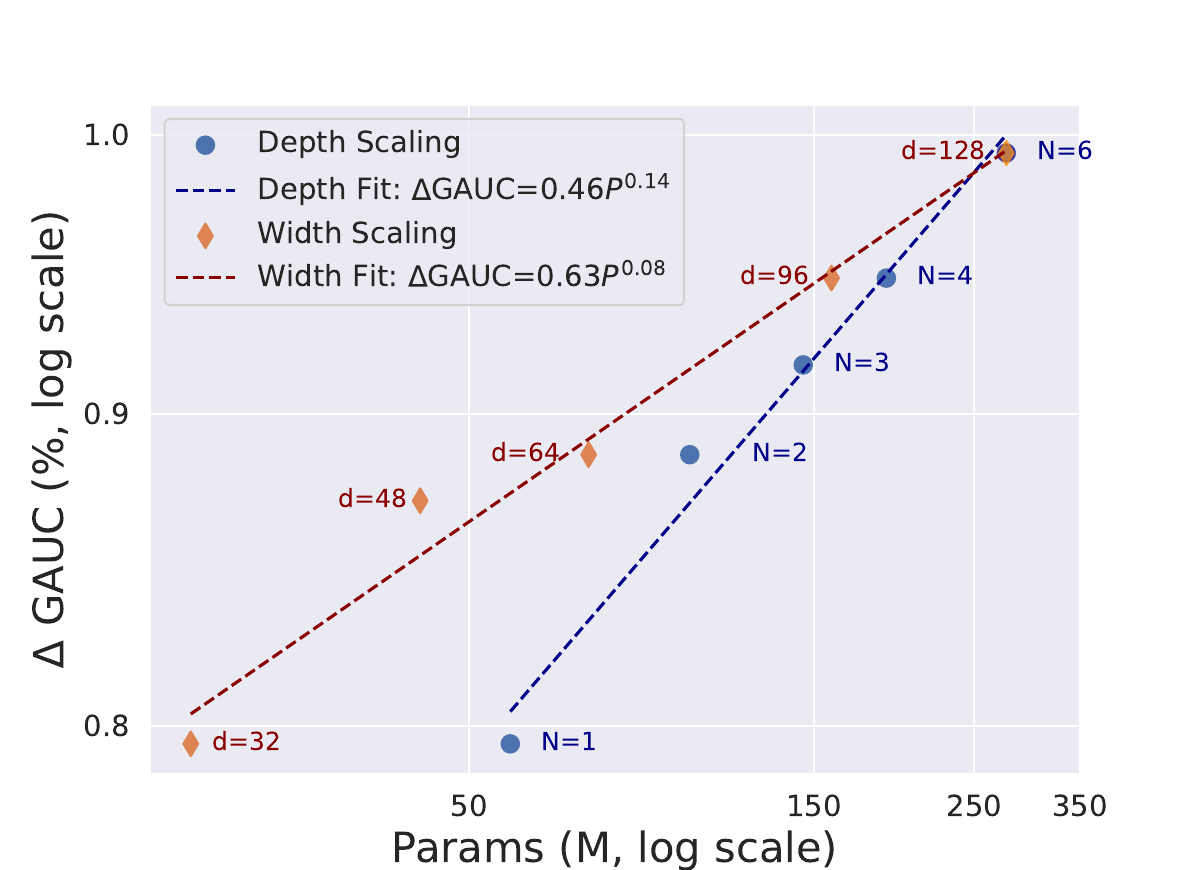}
    \caption{The power-law relationship between model capacity and GAUC. As the model depth increases, the best-fit curve: $\Delta\text{GAUC} = 0.46 \times P^{0.14}$. As the model width increases, the best-fit curve: $\Delta\text{GAUC} = 0.63 \times P^{0.08}$.}
    \label{fig: scaling_params}
\end{figure}

\subsection{Scaling Law (RQ3)}
We examine whether {\modelname} exhibits a predictable scaling behavior along two axes: (1) \textbf{computation overhead} $C$  (inference GFLOPs) and (2) \textbf{model capacity} $P$ (dense model parameters).
We follow \citet{kaplan2020scaling} and fit the $\Delta$GAUC to a power-law function $\Delta\text{GAUC}(X) = E \times X^{\alpha}$.
$X$ denotes the computation overhead $C$ or the model capacity $P$, while $E$ and $\alpha$ are coefficients to be fitted.

We adjust the \textbf{depth} $S$ (number of stacked blocks) and \textbf{width} $d$ (token dimension), with the inference GFLOPs ranging from $5$ to $50$ and the model capacity ranging from 30M to 0.3B.
After the log transformation, we present the results and the fitting curve in \Cref{fig: scaling_flops} and \Cref{fig: scaling_params}.
We observe that for both computational overhead and model capacity, the $\Delta$GAUC improves monotonically, following a power-law trend.
The growth trend is steeper when increasing model depth than width.
This suggests that by scaling depth, the model extracts intricate interactions.

\begin{table}[h]
\caption{Online A/B Results}
\vspace{-5pt}
\label{tab:ab}
\begin{tabular}{ccc}
\toprule
\textbf{Scenario} & \textbf{CTR} & \textbf{RPM} \\ \midrule
\emph{Guess}    & $+1.22\%$       & $+3.27\%$       \\
\emph{Post}     &   $+2.01\%$    & $+2.66\%$       \\ \bottomrule
\end{tabular}%
\vspace{-10pt}
\end{table}

\subsection{Online A/B Testing (RQ4)} 
We deployed EST in the Taobao display advertising, one of the largest e-commerce recommendation systems in the world, to evaluate its real business impact through online A/B testing. The experiments were conducted within \emph{QuanZhanTui}~\citep{DBLP:conf/kdd/00010LLB0D0Y0025}, an all-site advertising product, across two major scenarios: ``Guess What You Like'' (\emph{Guess}) and ``Post-Purchase'' (\emph{Post}).
The production baseline is a sophisticated MLP-based model enhanced with several state-of-the-art modules, including feature interaction like CAN~\citep{bian2020can} and SENET~\citep{hu2018squeeze}, sequence modeling like SIM~\citep{pi2020search} and MUSE~\citep{wu2025muse}, and content modeling like SimTier and MAKE~\citep{sheng2024enhancing}. As in~\Cref{tab:ab}, EST achieves substantial performance gains in both scenarios. Specifically, in the \emph{Guess} scenario, EST improves Click-Through Rate (CTR) by $\mathbf{1.22\%}$ and Revenue Per Mile (RPM) by $\mathbf{3.27\%}$, while in the \emph{Post} scenario, it raises CTR by $\mathbf{2.01\%}$ and RPM by $\mathbf{2.66\%}$.

%% file: sections/conclusion.tex

\section{Conclusion}

This paper tackles the challenge of adapting LLM-inspired scaling laws to industrial CTR prediction under strict latency constraints. By analyzing intrinsic differences between CTR inputs and LLM tokens in information density and modality, we derive two actionable design insights for efficient transformers. Guided by these insights, we propose EST, an efficiently scalable transformer architecture that unifies heterogeneous inputs within a single token sequence. EST introduces LCA to focus on the most informative interactions, which reduces redundant computation. It also adopts CSA to utilize content similarity for sparse modeling of long behaviors with little overhead. Offline evaluations confirm EST achieves state-of-the-art performance with a clear power-law scaling trend relative to model capacity and compute cost. Online A/B tests further validate its real-world impact, delivering significant gains of $3.27\%$ in RPM and $1.22\%$ in CTR.

%% file: sections/acknowledge.tex
\section*{Acknowledgement}
This work was sponsored by CCF-ALIMAMA TECH Kangaroo Fund (CCF-ALIMAMA OF 2025001).

%% file: sections/appendix.tex
\begin{center}
  {\bf \LARGE Supplementary Material}
\end{center}

\input{sections/related}